\documentclass[conference]{IEEEtran}
\IEEEoverridecommandlockouts
\usepackage{amsmath,amssymb,amsfonts}
\usepackage{algorithmic}
\usepackage{graphicx}
\usepackage{textcomp}
\usepackage{xcolor}
\usepackage{makecell}
\usepackage{hyperref}
\hypersetup{
    colorlinks=false,
    linkcolor=blue,
    filecolor=magenta,      
    urlcolor=cyan,
    pdftitle={Classification of IED-free EEG Responses for
Assisted Epilepsy Diagnosis},
    pdfpagemode=UseOutlines,
}
\usepackage{cleveref}
\usepackage{booktabs}
\usepackage[natbib=true, bibstyle=ieee,citestyle=numeric-comp]{biblatex}%
\usepackage{authblk}
\usepackage{lipsum}

\let\svthefootnote\thefootnote
\newcommand\freefootnote[1]{%
  \let\thefootnote\relax%
  \footnotetext{#1}%
  \let\thefootnote\svthefootnote%
}

\usepackage{setspace}
\begin{document}

\title{Classification of IED-free EEG Responses for Assisted Epilepsy Diagnosis}
\author[1]{Giacomo Zanardini}
\author[1]{Ryan Moesman}
\author[1,2]{Paul van der Kleij}
\author[2]{Robert van den Berg}
\author[1,$\ddagger$]{Justin Dauwels}
\affil[1]{Signal Processing Systems, Delft University of Technology, Delft, The Netherlands}
\affil[2]{Department of Neurology, Erasmus Medical Center, Rotterdam, The Netherlands}
\affil[$\ddagger$]{E-Mail: \textit{J.H.G.Dauwels}@tudelft.nl }

\maketitle

\begin{abstract}
Diagnosing epilepsy is challenging when routine EEGs lack interictal epileptiform discharges (IEDs). Intermittent photic stimulation (IPS) and hyperventilation (HV) can increase diagnostic yield, but their interpretation is subjective. We propose a reproducible pipeline that classifies EEG recordings acquired during stimulation procedures, using machine-learning features spanning temporal, spectral, wavelet, and connectivity domains, and a stacked ensemble to combine complementary feature sets. Performance is evaluated with leave-one-subject-out (LOSO) cross-validation on the TUH Epilepsy Corpus and a clinical Erasmus MC (EMC) cohort, including IED-free analyses on TUH. On TUH, ensembles achieve up to 97.8\% AUC / 93.1\% BAC on IED-free resting-state EEG and 94.1\% AUC / 86.8\% BAC on IED-free IPS. On EMC, IPS provides the strongest discrimination (79.4\% AUC / 73.9\% BAC), while HV performance benefits from stratifying subjects by responsiveness. These results indicate that stimulation-evoked activity, particularly IPS, contains meaningful discriminative information for IED-free epilepsy classification and that multi-domain ensembling improves robustness.

\end{abstract}


\section{Introduction}
\noindent Epilepsy is a chronic neurological disorder characterized by recurrent unprovoked seizures and affects roughly 50 million people worldwide, substantially impacting quality of life~\cite{noauthor_epilepsy_nodate}. The International League Against Epilepsy (ILAE) defines epilepsy as either two unprovoked seizures more than 24 hours apart or a $\geq$60\% risk of recurrence after a single unprovoked seizure~\cite{hirsch_ilae_2022,fisher_ilae_2014}. Routine diagnosis relies heavily on scalp EEG to detect seizures and interictal epileptiform discharges (IEDs)~\cite{fisher_epileptic_2005,pillai_interictal_2006}, yet IEDs are often absent and their interpretation can be challenging~\cite{basiri_focal_2019,noauthor_uc_nodate}. Intermittent photic stimulation (IPS) is routinely added to EEG to probe photosensitivity: stroboscopic flashes can elicit photo-paroxysmal responses that strongly support an epilepsy diagnosis when baseline EEG is inconclusive \cite{jayakar_clinical_1990,kasteleijnnolst_trenite_methodology_2012}. Similarly, hyperventilation (HV) reduces arterial \(\mathrm{CO}_2\) (hypocapnia), which in turn decreases cerebral blood flow and can produce a characteristic slowing of the EEG~\cite{holmes_does_2004}. In epilepsy, HV can elicit more pronounced slowing than in controls and may also trigger IEDs~\cite{guaranha_hyperventilation_2005}.
Even with IPS, HV, and a second sleep-deprived EEG~\cite{dellaquila_sleep_2022,noauthor_epilepsie_nodate}, many patients after a first seizure do not meet criteria for a definitive diagnosis and remain under a “wait-and-see” policy~\cite{epstein_facing_2021,berg_risk_2008}, prolonging diagnostic uncertainty and delaying treatment decisions. 

Recent research has increasingly applied machine learning (ML) and deep learning to EEG, predominantly targeting automatic seizure detection or interictal epileptiform discharge (IED) detection~\cite{carvajal-dossman_retraining_2025,shin_using_2023,tjepkemacloostermans_expert_2025,zou_accuracy_2024,khan_attention-enhanced_2025,wong_channel-annotated_2025}. Many approaches are developed on curated datasets that do not reflect routine clinical EEG complexity~\cite{zendehbad_systematic_2025,acharya_automated_2013,wong_eeg_2023,myers_diagnosing_2025}. Only a limited number of studies address patient-level epilepsy classification from interictal or IED-free EEG. Thangavel et al.~\cite{thangavel_improving_2022} showed that diagnostic information can be extracted from EEG without visible IEDs, and Mirwani and van der Kleij~\cite{mirwani_automated_2024,van_der_kleij_using_2025} extended these findings to large cohorts. However, these works focus mainly on resting-state EEG and provide little insight into the diagnostic contribution of routine activation procedures.

Despite the widespread use of intermittent photic stimulation (IPS) and hyperventilation (HV), their value for ML-based epilepsy diagnosis has been scarcely studied. To our knowledge, no prior work has presented an end-to-end ML pipeline explicitly targeting stimulation-evoked EEG segments (IPS/HV) for patient-level epilepsy classification, particularly in the absence of overt IEDs.

Here, we investigate whether stimulation-evoked EEG contains discriminative patterns that support automated epilepsy diagnosis. Using IPS recordings from the Temple University Hospital (TUH) Epilepsy Corpus~\cite{obeid_temple_2016} and an independent Erasmus MC cohort, we build on earlier interictal pipelines~\cite{thangavel_improving_2022,mirwani_automated_2024,van_der_kleij_using_2025} and develop, to the best of our knowledge, the first dedicated ML pipeline for IPS/HV-based epilepsy classification. We extract features capturing spectral, temporal, and spatial response characteristics and train boosted-tree and ensemble classifiers. Our goal is to quantify the added diagnostic value of IPS/HV and improve the clinical reliability of data-driven predictions.

\section{Datasets}
\noindent We used two complementary EEG datasets: the publicly available Temple University Hospital (TUH) Epilepsy Corpus and a clinically curated cohort from Erasmus MC (EMC). Both datasets were recorded with the international 10--20 system~\cite{obeid_temple_2016} and IPS periods. Hyperventilation is present only in EMC recordings. \cref{table:datasets_summary} summarizes the number of subjects (and recordings) in each dataset.
\begin{table}[ht]
\centering
\caption{Dataset composition: subject counts (epileptic vs. non-epileptic) for TUH and EMC. }
\begin{tabular}{lcc}
\toprule
Dataset & TUH & EMC \\
\midrule
Epileptic      & 13  & 40  \\
Non-Epileptic  & 18  & 101 \\
\midrule
Total          & 31  & 141 \\
\bottomrule
\end{tabular}\label{table:datasets_summary}\vspace{-0.3cm}
\end{table}
\subsection{TUH Epilepsy Corpus}
\noindent The Temple University Hospital (TUH) Epilepsy corpus~\cite{obeid_temple_2016} is the largest public collection of routine clinical EEG recordings. We selected recordings containing IPS, based on the presence of a \texttt{Photic~PH} trigger channel. This yielded 40 recordings from 31 subjects (13 epileptic, 18 non-epileptic), aged 12--88 years (12 females, 19 males), recorded between 2003 and 2011~\cite{obeid_temple_2016}.
The IPS protocol consisted of trains of flashes at increasing frequencies. Using the spectrogram of the \texttt{Photic~PH} channel, photic trains were identified as bursts whose spectrotemporal content followed the stimulus frequency sweep. Frequencies were typically applied in 2\,Hz steps from 1\,Hz to 21\,Hz, with occasional extensions to 23\,Hz. To construct the IED-free TUH subset, we used the IED annotations provided by Thangavel et al.~\cite{thangavel_improving_2022} and excluded recordings marked as containing IEDs. No additional manual IED review was performed in this study.

\subsection{Erasmus MC Dataset}
The EMC dataset comprises 141 adult subjects who presented to the emergency room after a first confirmed generalized tonic-clonic seizure and underwent routine EEG with IPS and HV at Erasmus MC, Rotterdam; Patients were excluded if they presented with a clear seizure aetiology (e.g. auto-immune encephalitis, stroke) or findings on MRI associated with a high recurrence risk, such as an intracranial tumor. All recordings were inspected by one of the authors and labelled as IED-free i.e. inconclusive as per ILAE guidelines. Based on at least one year of clinical follow-up, 40 subjects were classified as epileptic and 101 as non-epileptic: patients with a recurrent seizure were labelled \texttt{epileptic}, whereas those who remained seizure-free were labelled \texttt{healthy}. Only the first EEG per subject was included in this study.
Recordings were retrieved from the clinical EEG archive and pseudonymised. The study was approved by the local Medical Ethics Review Committee (MEC-2021-0145).

\section{Methods}
\subsection{Preprocessing}
\noindent All EEG recordings from both datasets were processed using a unified, fully automated pipeline. Channels not belonging to the international 10--20 system were discarded and all signals were converted to microvolts ($\mu$V) to enforce unit consistency across recordings.
To suppress power-line interference, we applied a notch filter at 50\,Hz for Erasmus MC (EMC) data and 60\,Hz for Temple University Hospital (TUH) data. A fourth-order zero-phase 1\,Hz high-pass Butterworth filter was then used to remove DC offsets and slow drifts. Segments containing extreme amplitude values (e.g. electrode pop or channel saturation) were removed: after segmenting the recordings into 1s windows using a sliding buffer, for each window, we computed the root mean square (RMS) per channel and rejected windows whose RMS exceeded a noise-based threshold, following the RMS-based artefact rejection strategy in~\cite{thomas_automated_2020}. This step removes residual high-amplitude artefacts while preserving physiologically plausible activity.  
After artifact exclusion, recordings were resampled to 200\,Hz (EMC) and 250\,Hz (TUH) to reduce data size and standardise temporal resolution for feature extraction. 
Finally resting state, photic stimulation, and hyperventilation segments were cropped for feature extraction.
\subsection{Feature Extraction}
\noindent From each cleaned EEG segment, we extract ten feature sets spanning temporal, spectral, time--frequency, and connectivity domains.
\paragraph*{Univariate Time Measures (UTM)}
Time-domain behaviour is captured using robust statistics, peak and zero-crossing counts, non-linear energy operators, signal energy measures, and Shannon entropy.
\paragraph*{Spectral features}
Relative power is computed within the standard $\delta$, $\theta$, $\alpha$, $\beta$, and $\gamma$ EEG frequency bands.
\paragraph*{Wavelet-based features}
Time--frequency representations are obtained using continuous and discrete wavelet transforms (CWT and DWT).
Both transforms are condensed into the mean square amplitude and the standard deviation of squared amplitudes of each channel’s coefficients, truncated to 13 scales for CWT and 6 decomposition levels for DWT.
\paragraph*{Stockwell-based features (mST and sST)}
Two feature sets are derived from the Stockwell transform: mST, which summarises band-power variability using the mean of the square root of the standard deviation, and sST, which captures distributional asymmetry via the skewness of the summed band power. All Statistics are computed across conventional EEG frequency bands.
\paragraph*{Connectivity features}
Inter-channel relationships are quantified using the maximum normalized cross-correlation (CC) and phase-locking value (PLV) across channel pairs and frequency bands.

\paragraph*{Graph features (GCC and GPLV)}
Connectivity matrices are represented as weighted graphs, from which global and nodal metrics are derived, including degree, strength, path length, efficiencies, clustering coefficients, centralities, assortativity, transitivity, neighbourhood overlap, and the matching index.

All feature types are extracted using multiple referential montages over window lengths ranging from 1s to 60s.
Statistical combiners are subsequently applied to summarise the extracted features, capturing both central tendency and variability.
The parameters used for feature extraction and combination are shown in \cref{tab:montages_segments_combiners}.

\begin{table}[ht]
    \centering
    \caption{Referential montages, segment lengths and statistical combiners used to summarize windowed features. }\label{tab:montages_segments_combiners}
    \begin{tabular}{ll}       
    \toprule
        \makecell[l]{Referential\\montages} & \makecell[l]{CAR, Cz, Laplacian, BipolarDB} \\
        \midrule
        \makecell[l]{Segment\\Lengths [s]} & \makecell[l]{1, 2, 5, 10, 20, 60}\\
        \midrule
        \makecell[l]{Statistical\\Combiners} & \makecell[l]{Mean, Median, Standard Deviation,\\Skewness, Kurtosis}\\
    \bottomrule
    \end{tabular}\vspace{-0.3cm}
\end{table}

\subsection{Classification \& Validation}
\noindent We formulate epilepsy diagnosis as a supervised binary classification problem and adopt Extreme Gradient Boosting (XGBoost)\cite{chen_xgboost_2016}. We use the following hyperparameters: \texttt{n\_estimators} = 100, \texttt{max\_depth} = 6, \texttt{subsample} = 0.9, \texttt{gamma} = 0.1, \texttt{learning\_rate} = 0.1, and a fold-specific \texttt{scale\_pos\_weight} to address class imbalance.

\paragraph{Feature set selection }
As a first stage, leave-one-subject-out cross-validation (LOSO-CV) is applied independently to each of the ten feature types. For every feature type we systematically evaluate all combinations of referential montage, segment length and statistical combiner. Each configuration is evaluated five times and average AUC is used to rank feature sets and to identify the optimal setting per feature type, which is  then retained for the ensemble stage.  

\paragraph{Stacking Classifier}
In the second stage, we build multi-feature ensembles by combining 2--10 of the best-scoring feature types. We fit a logistic meta-classifier that learns weights $\mathbf{w}\in\mathbb{R}^K$ on the simplex ($w_k\!\ge\!0$, $\sum_k w_k\!=\!1$) by solving \cref{eq:stacking}, with a penalty $\lambda=-\alpha\sum_{k=1}^{K}\log w_k$ (default $\alpha=0.05$) to prevent collapsing weights.

\begin{equation}\label{eq:stacking}
    \begin{aligned}
\min_{\mathbf{w}}\quad &
   -\frac{1}{N}\sum_{i=1}^{N}
     \left[y_i \log\sigma(z_i) +(1-y_i)\log\sigma(1-z_i)\right]+\lambda  \\[4pt]
\text{s.t.}\quad &
   \sum_{k=1}^{K} w_k = 1, \quad
   w_k \ge 0 \;\; \forall\,k, 
\end{aligned}
\end{equation}
where \( z_i = \sum_{k=1}^{K} w_k\,p_{ik},\) is the weighted log-odd for $i$t out $N$ training samples, $K$ the number of base classifiers, $p_{ik}$ the log-odds from classifier $k$ on the $i$th sample, and $\sigma(\cdot)$ is the logistic function.

\paragraph{Cross-validation}
To approximate clinical deployment where predictions must generalise to unseen patients, at every step we use LOSO CV. In each of $N$ folds, data from one subject is held out for testing, while the model is trained on the remaining $N\!-\!1$ subjects and evaluated on the left-out subject. Predictions from all folds are then aggregated to compute performance metrics.

In the ensembling step, each LOSO fold proceeds as follows: one subject is held out as test, and the remaining $N\!-\!1$ subjects are split into 70\% training and 30\% validation at the subject level. An instance of each feature-specific model is trained on the previously selected feature sets using only the training subjects. On the validation subjects, we collect the predicted probabilities from all base classifiers and fit the stacking classifier on the validation log-odds; during this step, we also determine the operating threshold that maximises the GMean score where \(\mathrm{GMean} = \sqrt{\mathrm{PPV} \cdot \mathrm{TPR}}\). The base models are then retrained on the union of training and validation subjects, the learned weights and threshold are fixed, and the ensemble is applied to the held-out subject. The process is repeated until every subject has served once as the test case. Every ensemble configuration is evaluated five times.

\subsection{Evaluation}
\noindent We evaluate (i) the EEG responses to stimulation and (ii) classifier performance using common classification metrics, and an ILAE-guided clinically relevant ROC region.
\paragraph{Stimulation Response}
To quantify the response to HV, we computed a spectral slowing index from band-power changes within the HV segment. EEG was segmented into 5\,s epochs CAR montage. To reduce transition artifacts, the HV segment was trimmed by 15\% at both the beginning and end. Within the trimmed segment, we defined (i) an early-HV baseline window consisting of the first 6 epochs ($\approx$30\,s) and (ii) a mid-to-late HV sample window consisting of 8 epochs centered at 75\% of the HV duration.

For each frequency band $b \in \{\delta, \theta, \alpha\}$, we computed the percent change in absolute power between baseline and sample windows:
\begin{equation}
\Delta_b = \frac{P_{\mathrm{sample},b} - P_{\mathrm{baseline},b}}{P_{\mathrm{baseline},b}} \times 100 .
\end{equation}
The slowing index was then defined as
\begin{equation}
S = -\Delta_{\alpha} + \Delta_{\theta} + \Delta_{\delta}.
\end{equation}
To enable cohort-level comparison, $S$ was standardized across subjects, and participants were classified as \emph{HV responders} (HV-R) if $S > 0$ and as \emph{HV non-responders} (HV-NR) otherwise. For validity, at least 3 epochs were required in each temporal window.

\paragraph{Metrics}
Performance is primarily evaluated using the Area Under the Receiver Operating Characteristic Curve (AUC) and Balanced Accuracy \(\mathrm{BAC} = \frac{\mathrm{Sens} + \mathrm{Spec}}{2}\) at \(\mathrm{Sens}=0.8\) to standardize the results. 
We define a \emph{clinically relevant} region in the ROC space, such that all classifiers within this region attain a reliable performance in clinical practice. This region lies above the line given by:
\begin{align}
    \mathrm{TPR}=\frac{P(\mathrm{posterior)}\cdot P(\mathrm{0})}{P(\mathrm{1})\cdot(1-P(\mathrm{posterior})}\cdot FPR
\end{align}
where \(P(\mathrm{posterior)}=0.6\) following ILAE guidelines~\cite{hirsch_ilae_2022}, where \(P(1)\) and \(P(0)\) denote the proportions of epileptic and healthy subjects in the evaluated cohort, respectively.

\section{Results}
\noindent This section reports classification performance on the TUH and EMC datasets across the evaluated segment types and feature configurations. We first present the best-performing single feature sets for each segment type, followed by the corresponding ensemble results. Performance is summarized using AUC and BAC, and comparisons are made against published reference values where available. Results are discussed by matching segment types to compare the proposed classification pipeline, and across segment types to asses the information content arising from stimulation procedures. 

\subsection{TUH dataset}
\begin{table}[ht]
\centering
\caption{TUH Dataset: best performing feature sets per segment type. * indicates the removal of recordings containing IEDs.}\label{tab:tuh_ss}
\begin{tabular}{llcc}
\toprule
\makecell[l]{\textbf{Segment}\\\textbf{Type}} & \textbf{Feature Set} & \makecell[c]{\textbf{AUC \%} } & \textbf{BAC \%}\\
\midrule
Resting State~\cite{mirwani_automated_2024}& Spectral & $87.1$ & $79.1$ \\
Resting State& UTM & $92.9 \pm 1.2$ & $85.5 \pm 2.1$ \\
\makecell[l]{Resting State*~\cite{mirwani_automated_2024}}& SST & $77.0$ & $71.6$ \\
\makecell[l]{Resting State*}& SST & $86.8 \pm 0.7$ & $86.9 \pm 2.6$ \\
IPS& UTM& $87.5 \pm 0.8$  & $81.9 \pm 2.2$ \\
\makecell[l]{IPS*} & Spectral & $84.5 \pm 1.2$  & $85.4 \pm 0$ \\
\bottomrule
\end{tabular}
\end{table}
The proposed pipeline achieves consistently higher performance than previously reported baselines when the same segment type is used (see~\cref{tab:tuh_ss}). For \textit{Resting State}, our best single-model configuration (UTM) reaches $92.9 \pm 1.2$ AUC and $85.5 \pm 2.1$ BAC, improving over the reference \textit{Resting State} result of $87.1$ AUC and $79.1$ BAC reported in~\cite{mirwani_automated_2024}. This indicates that, under matched segment types, the proposed feature extraction and modelling choices yield a clearer separation between classes.
When recordings containing IEDs are removed (\textit{Resting State*}), the performance gap becomes even more pronounced: our \textit{Resting State*} model attains $86.8 \pm 0.7$ AUC and $86.9 \pm 2.6$ BAC, compared to $77.0$ AUC and $71.6$ BAC in~\cite{mirwani_automated_2024}. Notably, the strong BAC for \textit{Resting State*} suggests that the proposed approach maintains good performance even under this stricter data regime.
For \textit{IPS}, the best single-model result (UTM) achieves $87.5 \pm 0.8$ AUC and $81.9 \pm 2.2$ BAC, demonstrating that IPS segments remain highly informative on TUH, though overall \textit{Resting State} still provides the strongest single-model discrimination. After removing IED recordings (\textit{IPS*}), the best configuration attains a lower AUC ($84.5 \pm 1.2$) but a higher BAC ($85.4 \pm 0$) comparable with the \textit{Resting State} segments.
\begin{figure}[ht]
    \centering
    \includegraphics[width=\linewidth]{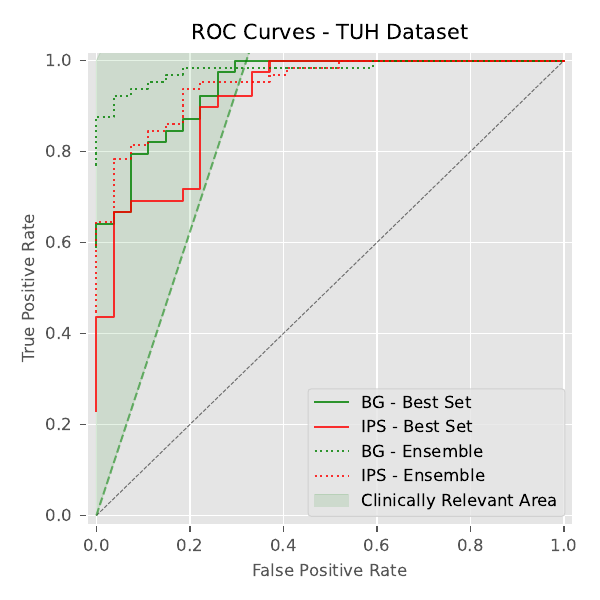}
    \vspace{-1cm}
    \caption{TUH dataset: ROC Curves for best single set and ensemble models.}
    \label{fig:tuh}
\end{figure}
\begin{table}[ht]
\centering
\caption{TUH dataset: best performing ensemble models per segment type. * indicates the removal of recordings containing IED}\label{tab:tuh_ens}
\begin{tabular}{llll}
\toprule
\makecell[l]{\textbf{Segment}\\\textbf{Type}} &  \makecell[l]{\textbf{Ensemble}\\\textbf{Size}} & \makecell[c]{\textbf{AUC \%} } & \textbf{BAC \%}\\
\midrule
Resting State & 3 & $97.3 \pm 1.4$ & $93.5 \pm 1.7$ \\
Resting State*~\cite{mirwani_automated_2024} & 3 & $76.0$ & $71.7$ \\
Resting State*~\cite{thangavel_improving_2022} & 3 & $71.5$ & $65.7$ \\
Resting State*  & 7 & $97.8 \pm 1.7$ & $93.1 \pm 3.4$ \\
IPS & 7 & $92.4 \pm 3.1$ & $89.1 \pm 4$ \\
IPS* & 7 & $94.1 \pm 3.1$ & $86.8 \pm 2.4$ \\
\bottomrule
\end{tabular}
\end{table}

Ensembling yields a substantial performance boost across TUH segment types, with the largest gains observed for \textit{Resting State}~(\cref{tab:tuh_ens}). Our \textit{Resting State} ensemble reaches $97.3 \pm 1.4$ AUC and $93.5 \pm 1.7$ BAC, and the \textit{Resting State*} ensemble reaches $97.8 \pm 1.7$ AUC and $93.1 \pm 3.4$ BAC. These results markedly exceed prior ensemble reference values on the same \textit{Resting State*} segment type (e.g., $76.0$ AUC / $71.7$ BAC in~\cite{mirwani_automated_2024} and $71.5$ AUC / $65.7$ BAC in~\cite{thangavel_improving_2022}), demonstrating that the proposed method scales effectively with ensemble aggregation and is not merely benefiting from segmentation choice.For IPS, ensembles also improve performance (up to $92.4 \pm 3.1$ AUC / $89.1 \pm 4$ BAC for IPS), though results remain slightly below \textit{Resting State} ensembles on TUH. Removing IED recordings increases AUC for IPS* ($94.1 \pm 3.1$) but reduces BAC ($86.8 \pm 2.4$). We observe (see~\cref{fig:tuh}) that both segment types attain operating points within the clinically relevant region. Overall, TUH results indicate that \textit{Resting State} segments are the most consistently high-performing segment type in this setting, while IPS segments provide competitive performance and may offer complementary information reflected in the ensemble gains.
\begin{figure}[ht]
    \centering
    \includegraphics[width=\linewidth]{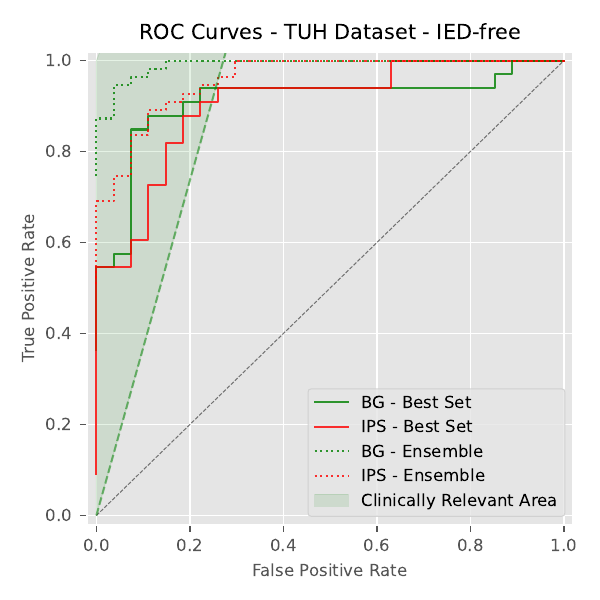}
    \vspace{-1cm}
    \caption{IED-free TUH dataset: ROC Curves for best single set and ensemble models.}
    \label{fig:tuh_noieds}
\end{figure}

\subsection{EMC Dataset}
On EMC single feature sets, performance depends more strongly on segment type~(\cref{tab:emc_ss}). For \textit{Resting State}, our best single-model result (GPLV; $69.1 \pm 1.2$ AUC, $62.2 \pm 3.5$ BAC) is lower than the reference \textit{Resting State} AUC values reported by~\cite{mirwani_automated_2024} ($72.0$) and~\cite{van_der_kleij_using_2025} ($71.4$), suggesting that resting-state connectivity features alone are less robust under the present evaluation pipeline. In contrast, the segment types introduced and evaluated in this work yield stronger discrimination: \textit{IPS} achieves the best single-model performance ($75.0 \pm 2.6$ AUC, $70.5 \pm 3.5$ BAC), outperforming both \textit{Resting State} and \textit{HV}. The \textit{HV} segment type performs moderately ($67.4 \pm 1.9$ AUC, $64.8 \pm 2.4$ BAC), indicating that, on EMC, the discriminative signal is more concentrated in IPS segments than in HV or resting-state segments.
\begin{table}[ht]
\centering
\caption{EMC Dataset: best performing feature sets per segment type.}\label{tab:emc_ss}
\begin{tabular}{llcc}
\toprule
\makecell[l]{\textbf{Segment}\\\textbf{Type}} & \textbf{Feature Set} & \makecell[c]{\textbf{AUC \%} } & \textbf{BAC \%}\\
\midrule
Resting State~\cite{mirwani_automated_2024}  & GPLV         & $72.0$ & $N/A$ \\
Resting State~\cite{van_der_kleij_using_2025}  & MST         & $71.4$ & $N/A$ \\
Resting State  & GPLV         & $69.1 \pm 1.2$ & $62.2 \pm 3.5$ \\
IPS & PLV       & $75.0 \pm 2.6$ & $70.5 \pm 3.5$ \\
HV  & PLV       & $67.4 \pm 1.9$ & $64.8 \pm 2.4$ \\
\bottomrule
\end{tabular}
\end{table}

Ensembling improves all EMC segment types and changes the relative standing of \textit{Resting State} compared to the literature (\cref{tab:emc_ensembles}). Our \textit{Resting State} ensemble reaches $75.6 \pm 4.4$ AUC and $68.3 \pm 5.9$ BAC, exceeding the \textit{Resting State} reference ensemble value of $70.0$ AUC and $67$ BAC in~\cite{mirwani_automated_2024}. Across the segment types evaluated in this work, \textit{IPS} remains the top performer on EMC: the IPS ensemble reaches $79.4 \pm 5.9$ AUC and $73.9 \pm 6.8$ BAC, outperforming the best \textit{Resting State} ensemble by $+3.8$ AUC and $+5.6$ BAC, and substantially outperforming \textit{HV}. The overall \textit{HV} ensemble achieves $72.3 \pm 8.2$ AUC and $64.6 \pm 3.7$ BAC, with high variance across folds. Splitting HV into responsive and non-responsive subsets clarifies this behaviour: \textit{HV-R} matches the HV AUC ($72.3 \pm 1.2$) but with markedly reduced variability, while \textit{HV--NR} collapses toward chance ($54.1 \pm 6.1$ AUC, $55.4 \pm 6.8$ BAC), indicating limited discriminative content in non-responsive HV periods.
\begin{figure}[ht]
    \centering
    \includegraphics[width=\linewidth]{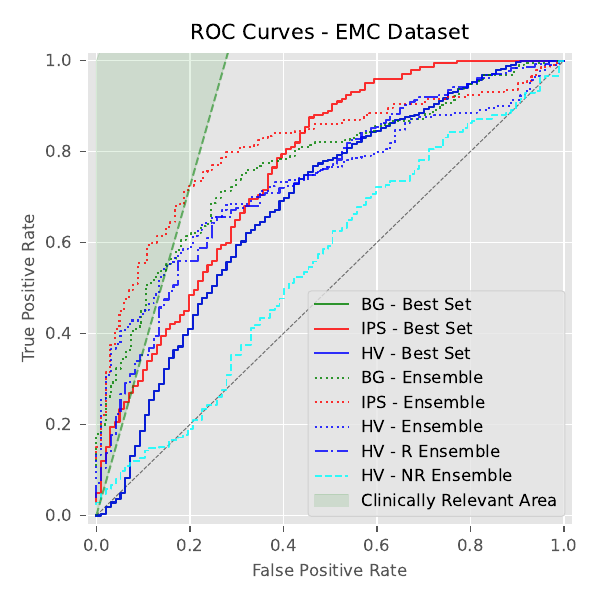}
    \vspace{-1cm}
    \caption{EMC dataset: ROC Curves for best single set and ensemble models.}
    \label{fig:emc}
\end{figure}The EMC results highlight that the segment types employed in this work, particularly \textit{IPS} and \textit{Resting State} provide the strongest performance, while \textit{HV} benefits substantially from ensembling but is not the dominant segment type on this dataset. We observe (see~\cref{fig:emc}) that most segment types attain operating points within the clinically relevant region, except the best individual set alone and the HV-NR ensemble. These results suggest that subjects exhibiting a physiological HV response constitute a more homogeneous subgroup with more discriminable EEG features. The same classifier achieves markedly higher epilepsy detection rates in HV responders, indicating that HV response status may serve as a useful stratification criterion in clinical practice.
\begin{table}[ht]
\centering
\caption{EMC Dataset: Best performing ensembles and reference values. R=Responders, NR=Non-Responders.}\label{tab:emc_ensembles}
\begin{tabular}{llll} 
\toprule
\makecell[l]{\textbf{Segment}\\\textbf{Type}} & \makecell[l]{\textbf{Ensemble}\\\textbf{Size}}& \makecell[c]{\textbf{AUC \%} } & \textbf{BAC \%}\\
\midrule
Resting State~\cite{mirwani_automated_2024} & 4 & $70.0$ & $67$ \\
Resting State& 5& $75.6 \pm 4.4$ & $68.3 \pm 5.9$ \\
IPS& 3& $79.4 \pm 5.9$ & $73.9 \pm 6.8$ \\
HV& 6& $72.3 \pm 8.2$ & $64.6 \pm 3.7$ \\
HV--R  & 3& $72.3 \pm 1.2$ & $63.4 \pm 2.4$ \\
HV--NR & 5& $54.1 \pm 6.1$ & $55.4 \pm 6.8$ \\
\bottomrule
\end{tabular}
\end{table}

\section{Discussion}
\noindent Across both datasets, the proposed pipeline shows that diagnostically relevant information can be extracted from routine EEG even when recordings lack visible interictal epileptiform discharges (IEDs), and that stimulation-evoked segments can be leveraged in a reproducible end-to-end machine-learning workflow. Relative to prior IED-free approaches that predominantly focus on resting-state EEG, this work explicitly targets routine activation procedures  (i.e. IPS, and HV in EMC), using multi-domain features (temporal, spectral, wavelet/time--frequency, and connectivity/graph measures) and patient-level leave-one-subject-out (LOSO) validation. The stacking step further provides a reliable combination of feature-specific models, supporting multi-feature aggregation rather than relying on a single feature family.

On the TUH dataset, performance improvements over previously reported baselines are consistent when matching segment type and evaluation setting, with ensembling providing the largest gains. In the stricter IED-free setting, the resting-state ensemble reaches $97.8\%$ AUC and $93.1\%$ BAC, substantially exceeding the referenced IED-free resting-state performance (e.g., $77.0\%$ AUC and $71.6\%$ BAC). Importantly for the focus of this work, IPS segments are also highly informative: in the IED-free IPS condition (IPS$^{\ast}$), the ensemble achieves $94.1\%$ AUC and $86.8\%$ BAC. In both resting-state and IPS, ROC operating points fall within the clinically relevant region defined in this study.

In EMC, which is IED-free by design and reflects a first-seizure cohort classified via follow-up, performance depends more strongly on segment type. Notably, IPS provides the strongest discrimination: the IPS ensemble achieves $79.4\%$ AUC and $73.9\%$ BAC, outperforming the best resting-state ensemble ($75.6\%$ AUC and $68.3\%$ BAC). HV performance shows substantial variance, which is clarified by stratifying subjects by HV response: HV responders retain $72.3\%$ AUC with reduced variability, while HV non-responders collapse toward chance ($54.1\%$ AUC and $55.4\%$ BAC). Consistent with the ROC analysis described in the Results, most EMC segment types achieve operating points within the clinically relevant region, except the best individual set alone and the HV--NR ensemble.

Interpretation of TUH performance with vs.\ without IED-containing recordings should remain cautious, as the IED-free TUH condition was obtained by excluding only two subjects (both epileptic). In addition, cross-cohort differences in patient populations and acquisition protocols likely contribute to differences between TUH and EMC performance.

Several limitations motivate clear next steps. While two datasets with different protocols were analysed, the study remains limited by the modest cohort size, particularly the TUH subset, and by the absence of broader multi-centre or leave-one-institution-out validation. Because HV was available only in EMC and the present analysis evaluated segment types separately, we did not quantify the additive predictive value of jointly combining IPS and HV; this remains an important direction for future work. In addition, HV performance may be affected by variability in clinical HV execution and patient compliance, which were not controlled prospectively in this retrospective dataset. Although the ensemble improves performance, its clinical interpretability remains limited because the present study does not analyse the relative contribution of individual features or physiological EEG characteristics driving each prediction. Finally, operating thresholds within the \textit{clinically relevant} region can be tuned to prioritise fewer false positives or higher sensitivity depending on clinical policy, as the proposed pipeline is intended to \emph{assist, not replace} clinicians in diagnostically ambiguous cases.


\section{Conclusion}
\noindent This study investigated automated epilepsy classification using EEG responses to routine activation procedures, demonstrating promising performance across two independent cohorts and showing that stimulation-evoked segments, particularly IPS, can carry clinically relevant discriminative information even in the absence of overt IEDs. On TUH, ensembling yields very strong performance (up to 97.3\% AUC/ 93.5\% BAC for resting state and 92.4\% AUC / 89.1\% BAC for IPS), including in the IED-free condition. On EMC, IPS provides the best results (IPS ensemble 79.4\% AUC / 73.9\% BAC), while HV performance benefits from stratification by physiological response. These findings support the proposed reproducible pipeline as a practical step towards more reliable data-driven support in diagnostically challenging EEG evaluations, while underscoring the need for broader multi-centre validation and robust handling of artifact-prone recordings.


%
\printbibliography

@inproceedings{chen_xgboost_2016,
    title = {{XGBoost}: {A} {Scalable} {Tree} {Boosting} {System}},
    shorttitle = {{XGBoost}},
    doi = {10.1145/2939672.2939785},
    booktitle = {Proceedings of the 22nd {ACM} {SIGKDD} {International} {Conference} on {Knowledge} {Discovery} and {Data} {Mining}},
    author = {Chen, Tianqi and Guestrin, Carlos},
    month = aug,
    year = {2016},
    note = {arXiv:1603.02754 [cs]},
    pages = {785--794},
}

@misc{noauthor_epilepsy_nodate,
	title = {Epilepsy},
	url = {https://www.who.int/news-room/fact-sheets/detail/epilepsy},
	language = {en},
}

@article{hirsch_ilae_2022,
    title = {{ILAE} definition of the {Idiopathic} {Generalized} {Epilepsy} {Syndromes}: {Position} statement by the {ILAE} {Task} {Force} on {Nosology} and {Definitions}},
    volume = {63},
    issn = {0013-9580, 1528-1167},
    shorttitle = {{ILAE} definition of the {Idiopathic} {Generalized} {Epilepsy} {Syndromes}},
    doi = {10.1111/epi.17236},
    language = {en},
    number = {6},
    journal = {Epilepsia},
    author = {Hirsch, Edouard and French, Jacqueline and Scheffer, Ingrid E. and Bogacz, Alicia and Alsaadi, Taoufik and Sperling, Michael R. and Abdulla, Fatema and Zuberi, Sameer M. and Trinka, Eugen and Specchio, Nicola and Somerville, Ernest and Samia, Pauline and Riney, Kate and Nabbout, Rima and Jain, Satish and Wilmshurst, Jo M. and Auvin, Stephane and Wiebe, Samuel and Perucca, Emilio and Moshé, Solomon L. and Tinuper, Paolo and Wirrell, Elaine C.},
    month = jun,
    year = {2022},
    pages = {1475--1499},
}

@article{fisher_ilae_2014,
    title = {{ILAE} {Official} {Report}: {A} practical clinical definition of epilepsy},
    volume = {55},
    copyright = {Wiley Periodicals, Inc. © 2014 International League Against Epilepsy},
    issn = {1528-1167},
    shorttitle = {{ILAE} {Official} {Report}},
    doi = {10.1111/epi.12550},
    language = {en},
    number = {4},
    journal = {Epilepsia},
    author = {Fisher, Robert S. and Acevedo, Carlos and Arzimanoglou, Alexis and Bogacz, Alicia and Cross, J. Helen and Elger, Christian E. and Engel Jr, Jerome and Forsgren, Lars and French, Jacqueline A. and Glynn, Mike and Hesdorffer, Dale C. and Lee, B.i. and Mathern, Gary W. and Moshé, Solomon L. and Perucca, Emilio and Scheffer, Ingrid E. and Tomson, Torbjörn and Watanabe, Masako and Wiebe, Samuel},
    year = {2014},
    pages = {475--482},
}

@article{fisher_epileptic_2005,
    title = {Epileptic {Seizures} and {Epilepsy}: {Definitions} {Proposed} by the {International} {League} {Against} {Epilepsy} ({ILAE}) and the {International} {Bureau} for {Epilepsy} ({IBE})},
    volume = {46},
    issn = {1528-1167},
    shorttitle = {Epileptic {Seizures} and {Epilepsy}},
    doi = {10.1111/j.0013-9580.2005.66104.x},
    language = {en},
    number = {4},
    journal = {Epilepsia},
    author = {Fisher, Robert S. and Boas, Walter van Emde and Blume, Warren and Elger, Christian and Genton, Pierre and Lee, Phillip and Engel Jr., Jerome},
    year = {2005},
    pages = {470--472},
}

@article{pillai_interictal_2006,
    title = {Interictal {EEG} and the {Diagnosis} of {Epilepsy}},
    volume = {47},
    issn = {1528-1167},
    doi = {10.1111/j.1528-1167.2006.00654.x},
    language = {en},
    number = {s1},
    journal = {Epilepsia},
    author = {Pillai, Jyoti and Sperling, Michael R.},
    year = {2006},
    pages = {14--22},
}

@article{basiri_focal_2019,
    title = {Focal epilepsy without interictal spikes on scalp {EEG}: {A} common finding of uncertain significance},
    author={Basiri, Reza and Shariatzadeh, Aidin and Wiebe, Samuel and Aghakhani,Yahya },
    volume = {150},
    issn = {0920-1211},
    shorttitle = {Focal epilepsy without interictal spikes on scalp {EEG}},
    doi = {10.1016/j.eplepsyres.2018.12.009},
    journal={Epilepsy Research}
}

@misc{noauthor_uc_nodate,
    title = {{UC} {Davis} {Department} of {Neurology} - {Epilepsy} {FAQs}},
    url = {https://health.ucdavis.edu/neurology/subspecialties/epilepsy_faqs.html}
}

@article{jayakar_clinical_1990,
    title = {Clinical correlations of photoparoxysmal responses},
    volume = {75},
    issn = {0013-4694},
    doi = {10.1016/0013-4694(90)90178-M},
    number = {3},
    journal = {Electroencephalography and Clinical Neurophysiology},
    author = {Jayakar, Prasanna and Chiappa, Keith H.},
    month = mar,
    year = {1990},
    pages = {251--254},
}

@article{kasteleijnnolst_trenite_methodology_2012,
    title = {Methodology of photic stimulation revisited: {Updated} {European} algorithm for visual stimulation in the {EEG} laboratory},
    volume = {53},
    copyright = {http://onlinelibrary.wiley.com/termsAndConditions\#vor},
    issn = {0013-9580, 1528-1167},
    shorttitle = {Methodology of photic stimulation revisited},
    doi = {10.1111/j.1528-1167.2011.03319.x},
        
    language = {en},
    number = {1},
    journal = {Epilepsia},
    author = {Kasteleijn‐Nolst Trenité, Dorothée and Rubboli, Guido and Hirsch, Edouard and Martins Da Silva, Antonio and Seri, Stefano and Wilkins, Arnold and Parra, Jaime and Covanis, Athanasios and Elia, Maurizio and Capovilla, Giuseppe and Stephani, Ulrich and Harding, Graham},
    month = jan,
    year = {2012},
    pages = {16--24},
}

@article{dellaquila_sleep_2022,
    title = {Sleep deprivation: a risk for epileptic seizures},
    volume = {15},
    issn = {1984-0659},
    shorttitle = {Sleep deprivation},
    doi = {10.5935/1984-0063.20220046},
    number = {2},
    journal = {Sleep Science},
    author = {Dell’Aquila, Jason Tyler and Soti, Varun},
    year = {2022},
    pmid = {35755914},
    pmcid = {PMC9210558},
    pages = {245--249},
}

@misc{noauthor_epilepsie_nodate,
    title = {Epilepsie},
    url = {https://richtlijnendatabase.nl/richtlijn/epilepsie/elektrofysiologisch_onderzoek_bij_epilepsie.html},
    language = {en},
}

@article{epstein_facing_2021,
    title = {Facing epistemic and complex uncertainty in serious illness: {The} role of mindfulness and shared mind},
    volume = {104},
    issn = {1873-5134},
    shorttitle = {Facing epistemic and complex uncertainty in serious illness},
    doi = {10.1016/j.pec.2021.07.030},
    language = {eng},
    number = {11},
    journal = {Patient Education and Counseling},
    author = {Epstein, Ronald M.},
    month = nov,
    year = {2021},
    pmid = {34334265},
    pages = {2635--2642},
}

@article{berg_risk_2008,
    title = {Risk of recurrence after a first unprovoked seizure},
    volume = {49},
    copyright = {© 2008 The Author(s)},
    issn = {1528-1167},
    doi = {10.1111/j.1528-1167.2008.01444.x},
    language = {en},
    number = {s1},
    journal = {Epilepsia},
    author = {Berg, Anne T.},
    year = {2008},
    pages = {13--18},
}

@article{carvajal-dossman_retraining_2025,
    title = {Retraining and evaluation of machine learning and deep learning models for seizure classification from {EEG} data},
    volume = {15},
    copyright = {2025 The Author(s)},
    issn = {2045-2322},
    doi = {10.1038/s41598-025-98389-y},
    language = {en},
    number = {1},
    journal = {Scientific Reports},
    author = {Carvajal-Dossman¹, Juan Pablo and Guio, Laura and García-Orjuela, Danilo and Guzmán-Porras, Jennifer J. and Garces, Kelly and Naranjo, Andres and Maradei-Anaya, Silvia Juliana and Duitama, Jorge},
    month = may,
    year = {2025},
    pages = {15345},
}

@article{shin_using_2023,
    title = {Using spectral and temporal filters with {EEG} signal to predict the temporal lobe epilepsy outcome after antiseizure medication via machine learning},
    volume = {13},
    copyright = {2023 The Author(s)},
    issn = {2045-2322},
    doi = {10.1038/s41598-023-49255-2},
    language = {en},
    number = {1},
    journal = {Scientific Reports},
    author = {Shin, Youmin and Hwang, Sungeun and Lee, Seung-Bo and Son, Hyoshin and Chu, Kon and Jung, Ki-Young and Lee, Sang Kun and Park, Kyung-Il and Kim, Young-Gon},
    month = dec,
    year = {2023},
    pages = {22532},
}

@article{tjepkemacloostermans_expert_2025,
    title = {Expert level of detection of interictal discharges with a deep neural network},
    volume = {66},
    issn = {0013-9580},
    doi = {10.1111/epi.18164},
 
    number = {1},
    journal = {Epilepsia},
    author = {Tjepkema‐Cloostermans, Marleen C. and Tannemaat, Martijn R. and Wieske, Luuk and van Rootselaar, Anne‐Fleur and Stunnenberg, Bas C. and Keijzer, Hanneke M. and Koelman, Johannes H. T. M. and Tromp, Selma C. and Dunca, Ioana and van der Star, Baukje J. and de Koning, Myrthe E. and van Putten, Michel J. A. M.},
    month = jan,
    year = {2025},
    pmid = {39530797},
    pmcid = {PMC11742546},
    pages = {184--194},
}

@article{zou_accuracy_2024,
    title = {Accuracy of {Machine} {Learning} in {Detecting} {Pediatric} {Epileptic} {Seizures}: {Systematic} {Review} and {Meta}-{Analysis}},
    volume = {26},
    shorttitle = {Accuracy of {Machine} {Learning} in {Detecting} {Pediatric} {Epileptic} {Seizures}},
    doi = {10.2196/55986},
    language = {EN},
    number = {1},
    journal = {Journal of Medical Internet Research},
    author = {Zou, Zhuan and Chen, Bin and Xiao, Dongqiong and Tang, Fajuan and Li, Xihong},
    month = dec,
    year = {2024},
    pages = {e55986},
}

@article{khan_attention-enhanced_2025,
    title = {An {Attention}-{Enhanced} {3D}-{CNN} {Framework} for {Spectrogram}-{Based} {EEG} {Analysis} in {Epilepsy} {Detection}},
    issn = {2169-3536},
    doi = {10.1109/ACCESS.2025.3574646},
    journal = {IEEE Access},
    author = {Khan, Ziaullah and Dayal, Aakanksha and Kim, Hee-Cheol},
    year = {2025},
    pages = {1--1},
}

@article{wong_channel-annotated_2025,
    title = {Channel-annotated deep learning for enhanced interpretability in {EEG}-based seizure detection},
    volume = {103},
    issn = {1746-8094},
    doi = {10.1016/j.bspc.2024.107484},
    journal = {Biomedical Signal Processing and Control},
    author = {Wong, Sheng and Simmons, Anj and Rivera-Villicana, Jessica and Barnett, Scott and Sivathamboo, Shobi and Perucca, Piero and Ge, Zongyuan and Kwan, Patrick and Kuhlmann, Levin and O’Brien, Terence J.},
    month = may,
    year = {2025},
    pages = {107484},
}

@article{zendehbad_systematic_2025,
    title = {A systematic review of artificial intelligence techniques based on electroencephalography analysis in the diagnosis of epilepsy disorders: {A} clinical perspective},
    volume = {215},
    issn = {0920-1211},
    shorttitle = {A systematic review of artificial intelligence techniques based on electroencephalography analysis in the diagnosis of epilepsy disorders},
    doi = {10.1016/j.eplepsyres.2025.107582},
    journal = {Epilepsy Research},
    author = {Zendehbad, Seyyed Ali and Razavi, Athena Sharifi and Tabrizi, Nasim and Sedaghat, Zahra},
    month = sep,
    year = {2025},
    pages = {107582},
}

@article{acharya_automated_2013,
    title = {Automated {EEG} analysis of epilepsy: {A} review},
    volume = {45},
    issn = {0950-7051},
    shorttitle = {Automated {EEG} analysis of epilepsy},
    doi = {10.1016/j.knosys.2013.02.014},
    journal = {Knowledge-Based Systems},
    author = {Acharya, U. Rajendra and Vinitha Sree, S. and Swapna, G. and Martis, Roshan Joy and Suri, Jasjit S.},
    month = jun,
    year = {2013},
    pages = {147--165},
}

@article{wong_eeg_2023,
    title = {{EEG} datasets for seizure detection and prediction— {A} review},
    volume = {8},
    issn = {2470-9239},
    doi = {10.1002/epi4.12704},
    number = {2},
    journal = {Epilepsia Open},
    author = {Wong, Sheng and Simmons, Anj and Rivera‐Villicana, Jessica and Barnett, Scott and Sivathamboo, Shobi and Perucca, Piero and Ge, Zongyuan and Kwan, Patrick and Kuhlmann, Levin and Vasa, Rajesh and Mouzakis, Kon and O'Brien, Terence J.},
    month = feb,
    year = {2023},
    pmid = {36740244},
    pmcid = {PMC10235576},
    pages = {252--267},
}

@article{myers_diagnosing_2025,
    title = {Diagnosing {Epilepsy} with {Normal} {Interictal} {EEG} {Using} {Dynamic} {Network} {Models}},
    volume = {97},
    copyright = {© 2025 The Author(s). Annals of Neurology published by Wiley Periodicals LLC on behalf of American Neurological Association.},
    issn = {1531-8249},
    doi = {10.1002/ana.27168},
    language = {en},
    number = {5},
    journal = {Annals of Neurology},
    year = {2025},
    pages = {907--918},
}

@article{thangavel_improving_2022,
    title = {Improving automated diagnosis of epilepsy from {EEGs} beyond {IEDs}},
    volume = {19},
    issn = {1741-2560, 1741-2552},
    doi = {10.1088/1741-2552/ac9c93},
    language = {en},
    number = {6},
    journal = {Journal of Neural Engineering},
    author = {Thangavel, Prasanth and Thomas, John and Sinha, Nishant and Peh, Wei Yan and Yuvaraj, Rajamanickam and Cash, Sydney S and Chaudhari, Rima and Karia, Sagar and Jing, Jin and Rathakrishnan, Rahul and Saini, Vinay and Shah, Nilesh and Srivastava, Rohit and Tan, Yee-Leng and Westover, Brandon and Dauwels, Justin},
    month = dec,
    year = {2022},
    pages = {066017},
}

@mastersthesis{mirwani_automated_2024,
    title = {Automated {Epilepsy} {Diagnosis} beyond {IEDs} by {Multimodal} {Features} and {Deep} {Learning}},
    url = {https://resolver.tudelft.nl/uuid:c829feac-3482-47a3-9c3e-2e27e89056c0},
    school = {TU Delft},
    author = {Mirwani, Yash},
    year = {2024},
}

@mastersthesis{van_der_kleij_using_2025,
    title = {Using machine learning models trained on {IED}-free {EEGs} to support epilepsy diagnosis},
    url = {https://repository.tudelft.nl/record/uuid:e89c0857-496b-40a4-9361-c5a94680b908},
    school = {TU Delft},
    author = {van der Kleij, P. A.},
    year = {2025},
}

@article{obeid_temple_2016,
    title = {The {Temple} {University} {Hospital} {EEG} {Data} {Corpus}},
    volume = {10},
    issn = {1662-453X},
    doi = {10.3389/fnins.2016.00196},
    language = {English},
    journal = {Frontiers in Neuroscience},
    author = {Obeid, Iyad and Picone, Joseph},
    month = may,
    year = {2016},
}

@article{thomas_automated_2020,
    title = {Automated {Detection} of {Interictal} {Epileptiform} {Discharges} from {Scalp} {Electroencephalograms} by {Convolutional} {Neural} {Networks}},
    volume = {30},
    issn = {0129-0657},
    doi = {10.1142/S0129065720500306},
  
    number = {11},
    journal = {International Journal of Neural Systems},
    author = {Thomas, John and Jin, Jing and Thangavel, Prasanth and Bagheri, Elham and Yuvaraj, Rajamanickam and Dauwels, Justin and Rathakrishnan, Rahul and Halford, Jonathan J. and Cash, Sydney S. and Westover, Brandon},
    month = nov,
    year = {2020},
    pages = {2050030},
}

@article{guaranha_hyperventilation_2005,
    title = {Hyperventilation {Revisited}: {Physiological} {Effects} and {Efficacy} on {Focal} {Seizure} {Activation} in the {Era} of {Video}-{EEG} {Monitoring}},
    volume = {46},
    issn = {1528-1167},
    shorttitle = {Hyperventilation {Revisited}},
    doi = {10.1111/j.0013-9580.2005.11104.x},
    language = {en},
    number = {1},
    journal = {Epilepsia},
    author = {Guaranha, Mirian S. B. and Garzon, Eliana and Buchpiguel, Carlos A. and Tazima, Sérgio and Yacubian, Elza M. T. and Sakamoto, Américo C.},
    year = {2005},
    pages = {69--75},
}

@article{holmes_does_2004,
    title = {Does {Hyperventilation} {Elicit} {Epileptic} {Seizures}?},
    volume = {45},
    issn = {1528-1167},
    doi = {10.1111/j.0013-9580.2004.63803.x},
    language = {en},
    number = {6},
    journal = {Epilepsia},
    author = {Holmes, Mark D. and Dewaraja, Asanka S. and Vanhatalo, Sampsa},
    year = {2004},
    pages = {618--620},
}

\end{document}